 \documentclass[final,5p,times,twocolumn]{elsarticle}

\usepackage[utf8]{inputenc} 
\usepackage[T1]{fontenc}

\usepackage[subtle]{savetrees}

\usepackage[colorlinks]{hyperref}
\hypersetup{colorlinks = true,linkcolor = blue,anchorcolor =red,citecolor = blue,filecolor = red,urlcolor = red,pdfauthor=author}

\usepackage{amsmath}

\usepackage{amssymb}
\usepackage{lipsum}

\journal{Nuclear Instruments and Methods in Physics Research Section B}

\begin{document}

\begin{frontmatter}



\title{Electron Emission Channeling for lattice location of radioactive isotopes in single crystals: Improvements from a Timepix3 quad detector and new PyFDD data analysis software}


\author[ieap]{Eric David-Bosne\corref{cor1}}
\cortext[cor1]{Corresponding author:}
\ead{eric.bosne@utef.cvut.cz}

\author[kuleuven]{Ângelo Costa}
\author[ieap]{Benedikt Bergmann}
\author[ctn]{João Guilherme Correia}
\author[ciceco]{Manuel Ribeiro da Silva}
\author[ieap,uwb]{Petr Burian} 
\author[ctn]{Ulrich Wahl}

\affiliation[ieap]{organization={Institute of Experimental and Applied Physics, Czech Technical University},
            addressline={Husova 240/5}, 
            postcode={110 00 Prague}, 
            country={Czech Republic}}

\affiliation[kuleuven]{organization={KU Leuven, Quantum Solid-State Physics},
            addressline={Celestijnenlaan 200D}, 
            postcode={3001 Leuven}, 
            country={Belgium}}

\affiliation[ctn]{organization={Centro de Ciências e Tecnologias Nucleares, Departamento de Engenharia e Ciências Nucleares, Instituto Superior Técnico, Universidade de Lisboa},
            addressline={Estrada Nacional 10 km 139,7}, 
            postcode={2695-066 Bobadela LRS}, 
            country={Portugal}}

\affiliation[ciceco]{organization={CICECO, Institute of Materials, Universidade de Aveiro},
            addressline={Campus Universitário de Santiago}, 
            postcode={3810-193 Aveiro}, 
            country={Portugal}}

\affiliation[uwb]{organization={Faculty of Electrical Engineering, University of West Bohemia Univerzitni},
            addressline={Univerzitni 2795/26}, 
            postcode={301 00 Pilsen}, 
            country={Czech Republic}}

\begin{abstract}
Electron Emission Channeling (EC) is a powerful technique for the investigation of the lattice location of radioactive isotopes implanted into single crystals. After implantation the isotopes occupy certain lattice locations in the crystal, which can in some cases be altered by annealing. Upon decay, the emission of a charged particle, typically a beta, may result in a channeling trajectory when its starting lattice location is aligned with major symmetry axes or planes of the crystal. By measuring the emission anisotropy in the direction of these axes for distinct annealing temperatures, the lattice location of the isotope can be determined with great precision and insightful information can be obtained on how annealing affects the occupied sites. This work reports on the installation of a Timepix3 quad detector and Katherine Gen2 readout in an experimental setup located at ISOLDE at CERN. The large increase in the number of pixels of the Timepix3, in comparison to previously used pad detectors, required more sophisticated tools for data treatment and fitting of channeling patterns. From this need, the PyFDD software was born. Its latest update features an intuitive graphical interface, with tools for noise masking, pattern visualization, simulations browsing, chi-square or maximum likelihood based fits and gamma background correction.
\end{abstract}




\end{frontmatter}



\section{Introduction to electron emission channeling}
\label{introduction}

Electron Emission Channeling (EC) \cite{hofsass_emission_1992} is a nuclear technique that utilizes the channeling phenomenon to determine the lattice site location of radioactive dopants in single crystals. Understanding the distribution of dopants in crystals is a fundamental step in the development of semiconductor devices as macroscopic properties of doped crystals depend on the lattice site occupied by their dopants. For example changes in dopant site may affect, magnetic, optical and electrical properties. Furthermore, understanding whether and how the dopant site can be manipulated through high-temperature annealing can help fine-tune and improve device manufacture.

Channeling is a phenomenon that occurs when a charged particle is steered into a stable direction by the Coulomb potential originating from the atomic strings and planes inside a crystal. This phenomenon, often described when a beam of particles hits a crystal in a direction closely aligned with a main crystal axis, can also occur when the origin of the charged particles is a radioactive decay from an implanted isotope. In this case, the high dependency of channeling on the incidence angle of an external beam, is translated into high sensitivity to the origin of the emitted particle and therefore the lattice location of implanted isotope.

This work reports on the EC experiments conducted at ISOLDE, the online radioisotope separator facility located at CERN in Switzerland~\cite{catherall_isolde_2017}. Here, EC starts with the ion implantation of a radioactive isotope of interest into a single crystal. Produced isotopes are accelerated and electromagnetically separated into a beam of 30\,keV to 60\,keV that is then directed to a sample where the isotopes are ion implanted using the setup described in Ref.~\cite{silva_versatile_2013}. The implanted unstable isotope serves as a site probe and as a dopant proxy for its stable counterpart used in industrial doping processes. The beta emission occurring at the lattice site of the isotopes produces an external particle flux whose angular anisotropy is highly correlated with the implantation site. The anisotropy of the beta particle emission from the crystal is measured around a set of major crystallographic directions by orienting these crystal axes towards a two-dimensional (2D) position sensitive detector, typically of size $3\times3\,cm^2$.

The detector covers an angular range of about 6\,° when placed at 30\,cm from the sample, or about 3\,° when placed at 60\,cm. The latter is used when an increase in angular resolution is desired. The angular resolution is of the order of 0.05° standard deviation for the 30\,cm distance from detector to sample. Each measurement produces a 2D pattern of the observed angular distribution of the emitted beta particles, which is then repeated for each chosen crystal axis. 

Retrieving quantitative information from channeling patterns requires a comparison with simulated anisotropies. These simulations are carried out with an in-house software designed for the simulation of electron channeling based on the many-beam algorithm \cite{hofsass_emission_1992, hofsass_emission_1991, wahl_advances_2000}. Simulations need take into account the depth profile of the implantation, which can be calculated through SRIM \cite{ziegler_srim_2010}, as it impacts dechanneling calculations. This is due to the fact that the implantation depth, average in the tens of nanometer, has similar values to the dechanneling length, that lies in the range of several to a few hundred nanometers, depending on decay energy, host crystal and temperature. Due to their demanding computational power, simulations are conducted prior to the data analysis and compiled into a library of expected angular yield patterns for each lattice site.

Once the desired collection of simulations is complete, the data can be fitted with a linear combination of simulated yield patterns for a set of corresponding lattice sites. Each lattice site occupancy fraction is optimized to provide the best representation of the measurement (see Equation.\,\ref{eq:linear_comb}). This task can be performed through the PyFDD software\,\cite{david-bosne_generalized_2020}, previously released as a python scripting library and now upgraded.

The current paper presents the latest developments in PyFDD that make the analysis of channeling patterns easier through a graphical interface, more accessible through a single file executable distribution, and more precise through the addition of a gamma background correction feature. Furthermore, it describes the latest improvements in the experimental setup using a Timepix3 quad detector.

\section{Position sensitive detectors - Timepix3 upgrade}
\label{detectors}

Detectors for routine emission channeling measurements need to satisfy a few requirements. First of all, they need to have beta counting capabilities over an area of at least 3\,cm by 3\,cm, with low dead time and count rate capabilities ideally above 5k\,counts per second. Good energy resolution is also a desired feature, although not always necessary, as it allows for the selection of decay particles of specific energies which can help reduce background from characteristic X-ray. Finally, a relevant feature which needs to be taken into account is the ease of use of the detection system. EC channeling measurements of a yield pattern often require less than 10\,min and it is therefore necessary to have a reliable and practical system that allows to handle the data, visualize the measurement and control experimental conditions on a time scale of minutes.

Currently, at ISOLDE there are measurement setups based on two detectors, pad and Timepix3. The pad detectors\,\cite{wahl_position-sensitive_2004} are of an older type of detector but still reliable and useful, despite not being produced any more. These detectors have a 300-500\,$\mu$m thick silicon sensor with $22 \times 22$ pixels (pads) of 1.3\,mm by 1.3\,mm. The count rate of a maximum of 5000 counts per second has proven sufficient for most measurements. However, the biggest advantage of these detectors is the many years of experience and proven reliability in both hardware and data treatment software. In the last years, more modern detectors from the Medipix collaboration, firstly Timepix\,\cite{david-bosne_generalized_2020} and now Timepix3 \cite{poikela_timepix3_2014}, have been installed and tested. Both of these detectors were used in a quad configuration, consisting of four chips with a total of $512 \times 512$ energy sensitive pixels of 55\,$\mu$m by 55\,$\mu$m, with a silicon sensor of 300\,$\mu$m thickness. The Timepix3 quad chip board was designed by IEAP (Institute of Experimental and Applied Physics in Prague) and ZČU (University of West Bohemia in Pilsen) and was placed in a custom made vacuum chamber with a 15\,°C water cooling circuit for stabilizing the detector electronics, with negligible effect on the improvement of the energy resolution. The installed Timepix3 detector was operated with a 3.5\,keV energy threshold and possesses an energy resolution of 1.5\,keV when measured with a 59.5\,keV gamma source of \textsuperscript{241}Am.

\begin{figure}[t]
    \centering
    \includegraphics[width=0.90\linewidth]{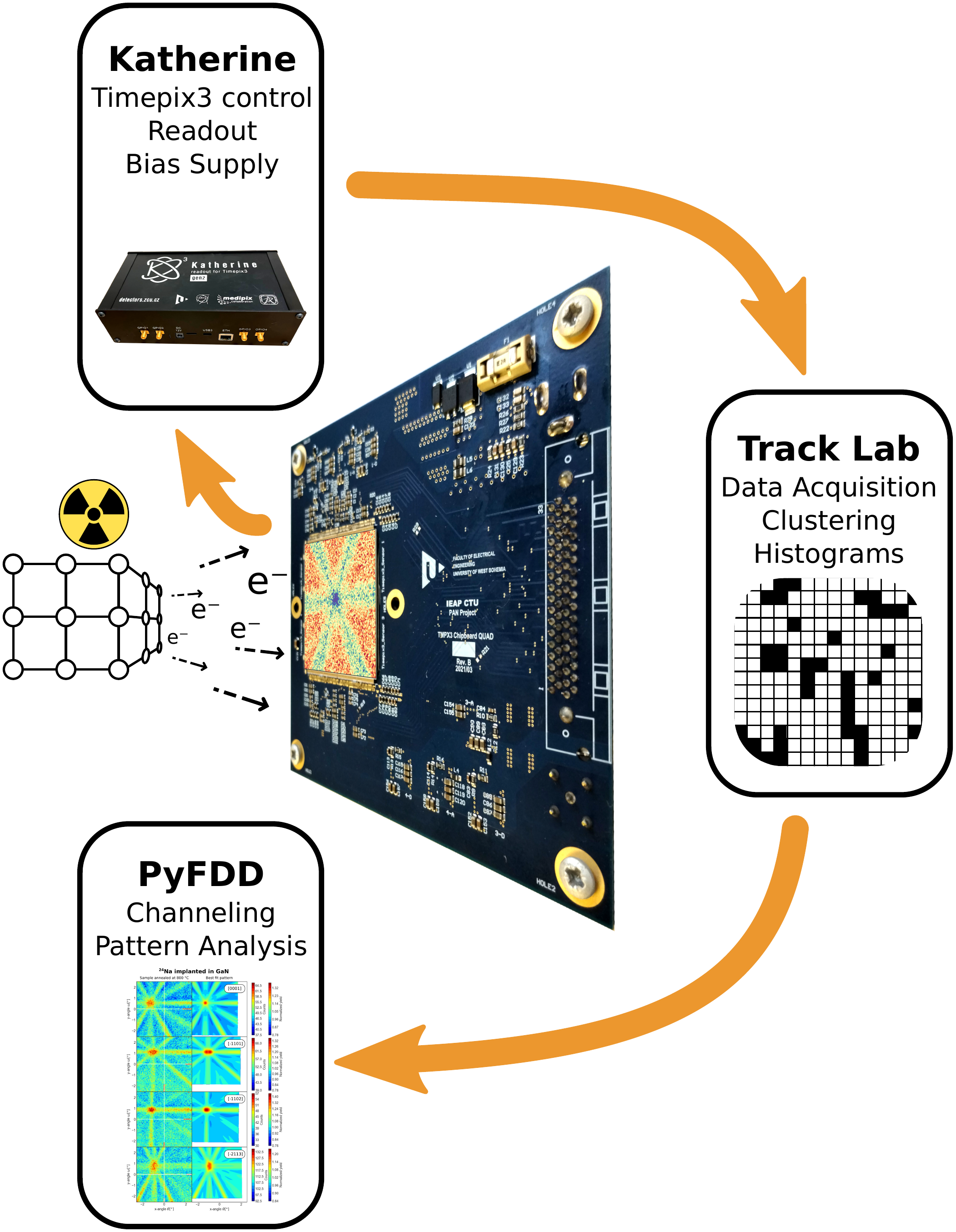}
    \caption{Schematic of data flow during a measurement. The Timepix3 quad detector captures the emission anisotropy from the beta particles channeling in the crystal lattice. A Katherine readout establishes the communication between the detector and a computer via Ethernet or USB. The Track Lab software receives this information, clusters the activated pixels into individual hits and builds a 2D pattern. PyFDD is then used to compare the acquired pattern with simulated yields for different lattice sites.}
    \label{fig:tpx3_scheme}
\end{figure}

The use of recently developed readout hardware, and control and data taking software, see Figure\,\ref{fig:tpx3_scheme}, has been a massive improvement in the rapid data acquisition and overall usability of the Timepix3 quad detectors. The readout, Katherine Gen2 \cite{burian_katherine_2017} allows for a readout speed of 16\,MHit/s by Ethernet or 40\,MHit/s by USB3 greatly surpassing the needs for EC and therefore removing any worries related to pile-up and data loss due to low communication bandwidth. The data taking software Track\,Lab \cite{manek_track_2024}, provides full control of the detector and data taking followed by high performance analysis and visualization of the data. Overall this enables to have a fast feedback on the ongoing experiment filling up the usability needs that would have been a drawback for the use of Timepix3 up to now.

The angular resolution of the measurement is given by,
\begin{equation}
\label{eq:angular_res}
\sigma_{\rm{Total}} = \arctan \left(\frac{\sqrt{(}\sigma_d^2+\sigma_b^2)}{d} \right)\quad,
\end{equation} 

\noindent where $\sigma_d$ is the contribution from detector position resolution, $\sigma_b$ is the contribution from the (1\,mm) implantation beam spot and $d$ is the distance between the detector and sample. From this formula, it can be understood that although an improvement in channeling pattern quality is achieved with the Timepix3, due to the way detector and beam contributions are added, the implantation beam spot is the limiting factor in further improving the overall angular resolution.

\section{PyFDD latest developments}

PyFDD\,\cite{david-bosne_generalized_2020} is a python tool, developed for fitting 2D channeling patterns with previously simulated channeling yield patterns. Its focus is on obtaining consistency in the analysis of data taken with PSD with either large (mm sized) or small ($\mu$m sized) pixels, as well as providing an accurate determination of statistical errors for the fit. PyFDD was initially developed as a python scripting software library, however, in the latest developments, it has been equipped with a user friendly PyQT interface and additional features such as including gamma background measurements to the fit function.  Fits with background are specially important in cases where the detected gamma background has a comparable intensity to the channeling beta particles and has a significant anisotropy over the detector area.

In the fit procedure, the use of a background pattern adds a contribution to the expected yield distribution pattern from simulations. From reference\,\cite{david-bosne_generalized_2020} the yield pattern, composed from a combination of lattice site simulations, $P_{\rm{sim}}$, is described as,

\begin{equation}
\label{eq:linear_comb}
\begin{aligned}
    P_{\rm{sim}}(f_1,f_2,f_3) = {} & f_1 P_1 + f_2 P_2 + f_3 P_3 + {}\\
    & (1 - f_1 -f_2 -f_3)\times P_{\rm{rand}}\quad,
\end{aligned}
\end{equation}

\noindent where $f_i$ is if the fraction of the occupied site which produces the respective channeling pattern $P_i$. Also, $P_{\rm{rand}}$ is a uniform (flat) pattern representing randomly distributed sites.

With this, a measured gamma background pattern $P_{\rm{\gamma\,bg}}$ can be added to the fit by first ensuring that the patterns have the same normalization, $\| P_{\rm{\gamma\,bg}} \| = \| P_{\rm{sim}} \|$. Note that, $P_{\rm{sim}} / \| P_{\rm{sim}} \|$ corresponds to the probability density function of the simulated yield pattern. The two normalized patterns can then be combined as,

\begin{equation}
\begin{aligned}
    P_{\rm{with\, bg}} = F_\gamma^{-1} P_{\rm{sim}} + (1 - F_\gamma^{-1}) P_{\rm{\gamma\,bg}} \quad.
\end{aligned}
\end{equation}

\noindent With $F_\gamma$ being the gamma background correction factor, defined as, 

\begin{equation}
\begin{aligned}
    F_{\gamma} = \frac{N_{\rm{total}}}{N_{\rm{total}} - N_{\gamma}} \quad ,
\end{aligned}
\end{equation}

\noindent where, $N_{\rm{total}}$ is the total number of counts in the measured pattern and $N_{\gamma}$ is the expected number of gammas detected during the measurement.

\begin{figure}
    \centering
    \includegraphics[width=0.95\linewidth]{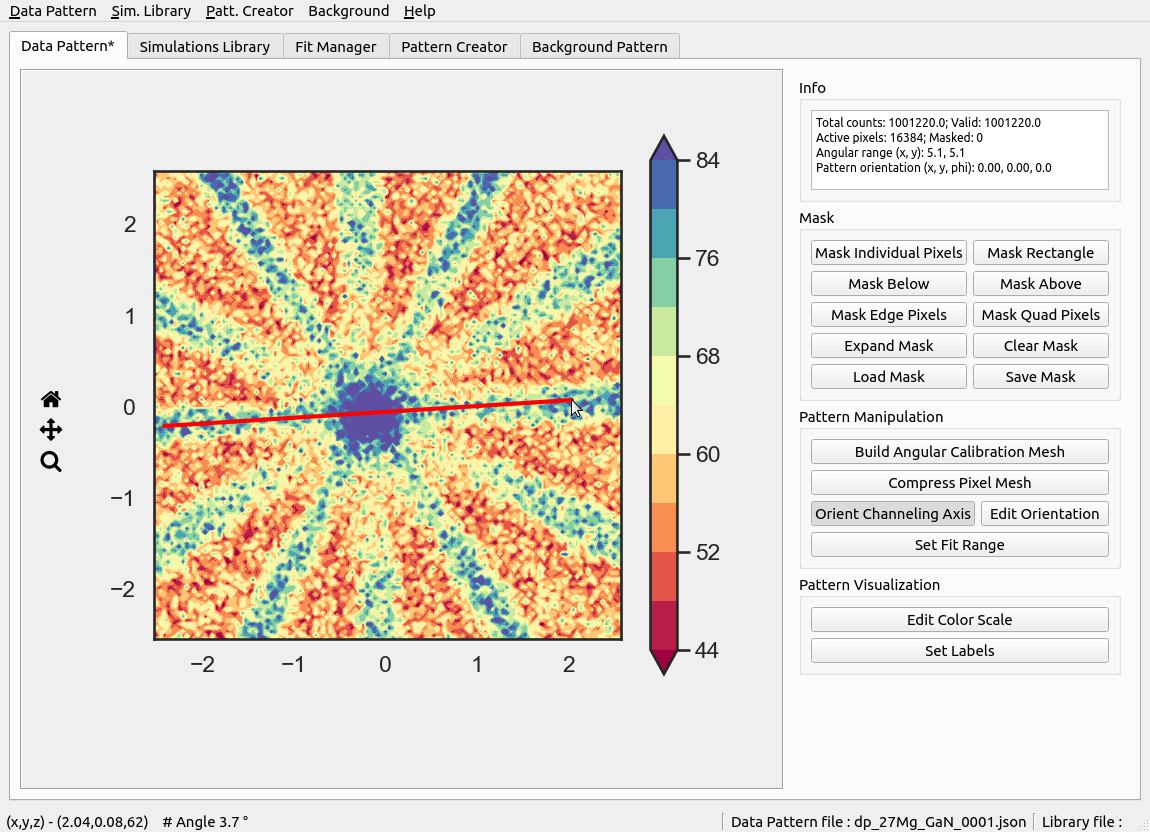}
    \caption{Screenshot of the PyFDD graphical interface with the Data Pattern tab selected. An example of a channeling pattern is being visualized on the app. The displayed pattern is a simulated representation of a GaN crystal observed along the [0001] axis for which the implanted beta emitter is located on a substitutional site. The red line on the pattern shows a user input for the axis center and rotation of the visualized pattern, later used during the fit process. }
    \label{fig:datapattern}
\end{figure}

The graphical interface of PyFDD includes 5 tabs each designed for a specific task. The first, named "Data Pattern" (see Figure\,\ref{fig:datapattern}), is responsible for visualizing measurement data and preparing it for analysis with tools such as pixel masking and indication of the center of the channeling effect. The second, named "Simulation Library", allows to browse a collection of channeling yield simulations for several lattice sites. The third, named "Fit manager", is responsible for running the fits with the desired lattice sites and optimization parameters. The fourth, named "Pattern Creator", is a tab for visualizing the expected pattern given the sites, fractions and other experimental conditions. This tab is useful for verifying an analysis or testing new experimental conditions. The fifth and last tab, named "Background Pattern", is used when including a gamma background measurement to the fit.

\section{Lattice location of \textsuperscript{121}Sn in diamond}

Emission channeling has a large record of contributions for direct observation of lattice site location in several crystallographic materials with various technological interests. One recent application of EC has been the determination of the lattice site location of dopant color centers in diamond. These color centers have recently attracted a lot of attention due to their possible use in quantum information processing applications. Specifically, EC has measured the existence of Sn split vacancy complexes in natural diamond\,\cite{wahl_direct_2020}. In the split vacancy complexes, the dopant is located in the bond centered (BC) site, between two carbon vacancies. Despite EC not detecting vacant sites, the nature of the occupied lattice site allows to determine if the complex exists.

In order to further demonstrate the applicability of the Timepix3 quad detector for EC, an experiment of \textsuperscript{121}Sn (116\,keV mean beta energy, $t_{1/2} = 27.06\,h$) implanted into artificial diamond grown by chemical vapour deposition (CVD) was performed. After implantation, this sample was annealed at 900\,°C. In this experiment, it was not possible to assess the gamma background as the setup with the Timepix3 detector does not have a shield that can be used to isolate the gamma from the beta radiation. For this reason, the occupancy values are here given as relative values instead of absolute fractions.

\begin{figure}
    \centering
    \includegraphics[width=0.95\linewidth]{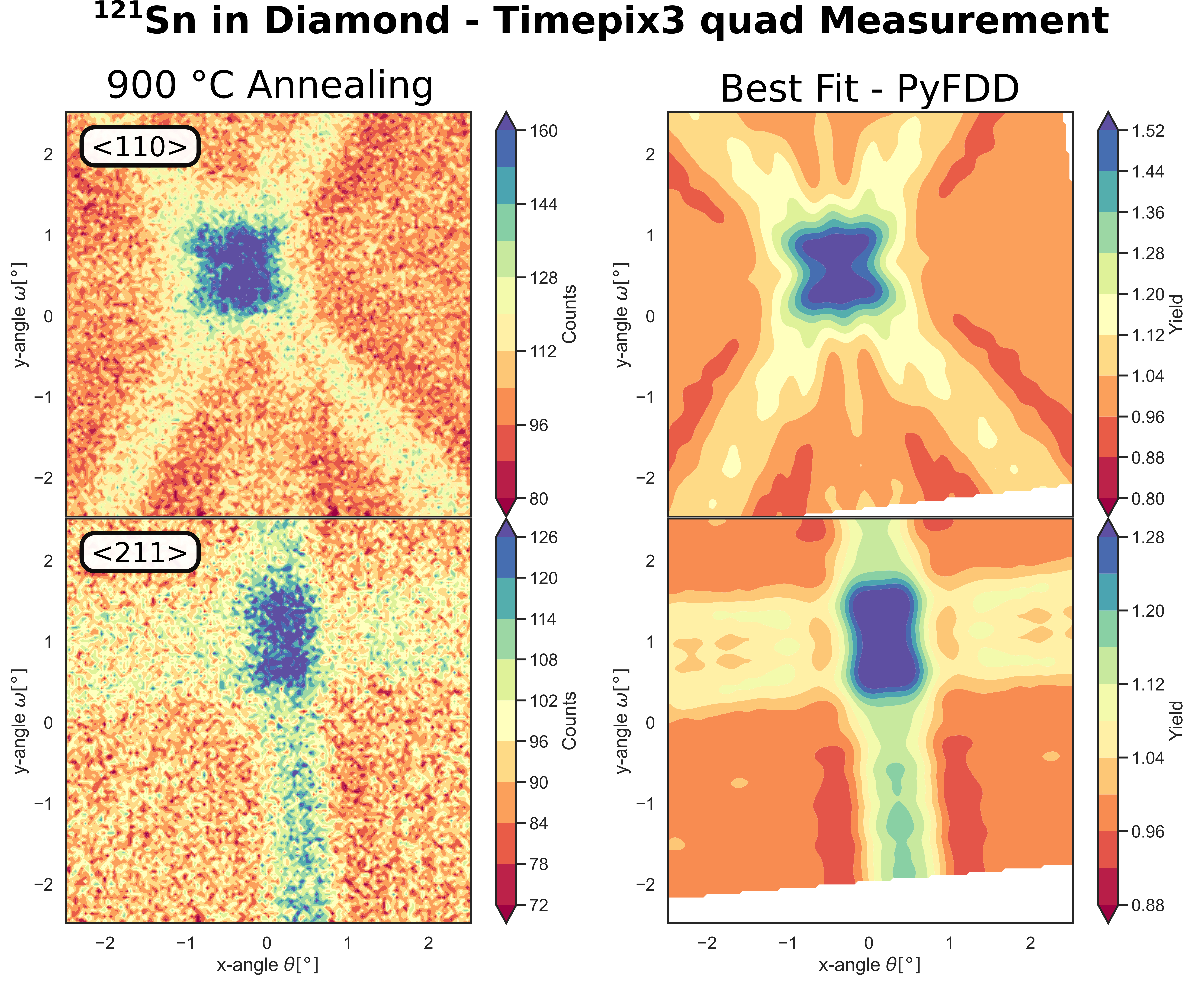}
    \caption{Measurements of \textsuperscript{121}Sn in diamond EC patterns taken with the Timepix3 Quad detector (on the left) and respective best fit patterns performed with PyFDD (on the right). The top two patterns correspond to the <110> direction, and the bottom two patterns to the <211> direction.}
    \label{fig:measurements}
\end{figure}

From the analysis of patterns taken for the <110> and <211> directions (see Figure\,\ref{fig:measurements}) it was concluded that the tin lattice site occupancy ratio between substitutional (S) and BC site is $1:0.67(3)$. This result differs slightly from the ratio $1:0.41$ reported in ref\,\cite{wahl_direct_2020}, which could indicate a somewhat different lattice site distribution in CVD diamond, or be due to the fact that a ten times higher implantation fluence was used in this experiment.

\section{Summary and conclusions}

Emission Channeling is a unique technique for lattice location due to wide elemental choice of probe atoms and ability to perform experiments at implantation fluences as low as $10^{11} \,\rm{cm}^{-2}$. The adoption of Timepix highly pixelized hybrid pixels detectors has improved the data quality of channeling patterns and motivated the development of new software for the analysis of channeling patterns with better angular resolution but lower number of counts per pixel.

Despite the higher capabilities of these modern detectors their full adaptation has so far been seriously slowed by the lack of fast, practical and user friendly software tools for data acquisition and analysis. The new Katherine readout and Track Lab software have solved these issues on the data acquisition side and now PyFDD, with a graphical user interface, is proving to be a useful tool for data analysis of 2D channeling patterns.

Although all these tools here described are applied for the case of EC, it is relevant to note that they can be just as well applied to Rutherford Backscattering Spectrometry with channeling \cite{david-bosne_use_2021} which is a more broadly available technique commonly used in many smaller accelerator laboratories.

\section*{Declaration of competing interest}
The authors declare that they have no known competing financial interests or personal relationships that could have appeared to influence the work reported in this paper.

\section*{Acknowledgements}
We appreciate the support of the ISOLDE Collaboration and technical teams. This work has received funding from KU~Leuven, \vfill\eject\noindent the Research Foundation Flanders (FWO, Belgium), and from the  Portuguese Foundation for Science and Technology (FCT), refs.: \href{https://doi.org/10.54499/UIDB/04349/2020}{UIDB/04349/2020}, \href{https://doi.org/10.54499/CERN/FIS-TEC/0003/2019}{CERN/FIS-TEC/0003/2019}, \href{https://doi.org/10.54499/CERN/FIS-TEC/0003/2021}{CERN/FIS-TEC/0003/2021}. The EU Horizon Europe Framework supported ISOLDE beam times through Grant Agreement 101057511 (EURO-LABS). E.~David-Bosne has been supported by the Global Postdoc Fellowship Program of the Czech Technical University in Prague. B.~Bergmann and E.~David-Bosne profited from funding of the Czech Science Foundation under grant number GM23-04869M.




\bibliographystyle{elsarticle-num} 
\bibliography{Channeling_2023}

\begin{thebibliography}{10}
\expandafter\ifx\csname url\endcsname\relax
  \def\url#1{\texttt{#1}}\fi
\expandafter\ifx\csname urlprefix\endcsname\relax\def\urlprefix{URL }\fi
\expandafter\ifx\csname href\endcsname\relax
  \def\href#1#2{#2} \def\path#1{#1}\fi

\bibitem{hofsass_emission_1992}
H.~Hofsäss, S.~Winter, S.~G. Jahn, U.~Wahl, {E. Recknagel}, Emission channeling studies in semiconductors, Nucl. Instr. and Meth. B 63~(1) (1992) 83--90.
\newblock \href {https://doi.org/10.1016/0168-583X(92)95174-P} {\path{doi:10.1016/0168-583X(92)95174-P}}.

\bibitem{catherall_isolde_2017}
R.~Catherall, W.~Andreazza, M.~Breitenfeldt, A.~Dorsival, G.~J. Focker, T.~P. Gharsa, T.~J. Giles, J.-L. Grenard, F.~Locci, P.~Martins, S.~Marzari, J.~Schipper, A.~Shornikov, T.~Stora, The {ISOLDE} facility, J. Phys. G: Nucl. Part. Phys. 44~(9) (2017) 094002.
\newblock \href {https://doi.org/10.1088/1361-6471/aa7eba} {\path{doi:10.1088/1361-6471/aa7eba}}.

\bibitem{silva_versatile_2013}
M.~R. Silva, U.~Wahl, J.~G. Correia, L.~M. Amorim, L.~M.~C. Pereira, A versatile apparatus for on-line emission channeling experiments, Rev. Sci. Instrum. 84~(7) (2013) 073506.
\newblock \href {https://doi.org/10.1063/1.4813266} {\path{doi:10.1063/1.4813266}}.

\bibitem{hofsass_emission_1991}
H.~Hofsäss, G.~Lindner, Emission channeling and blocking, Phys. Rep. 201~(3) (1991) 121--183.
\newblock \href {https://doi.org/10.1016/0370-1573(91)90121-2} {\path{doi:10.1016/0370-1573(91)90121-2}}.

\bibitem{wahl_advances_2000}
U.~Wahl, Advances in electron emission channeling measurements in semiconductors, Hyperfine Interac. 129~(1) (2000) 349--370.
\newblock \href {https://doi.org/10.1023/A:1012697429920} {\path{doi:10.1023/A:1012697429920}}.

\bibitem{ziegler_srim_2010}
J.~F. Ziegler, M.~D. Ziegler, J.~P. Biersack, {SRIM} – {The} stopping and range of ions in matter (2010), Nucl. Instr. and Meth. B 268~(11) (2010) 1818--1823.
\newblock \href {https://doi.org/10.1016/j.nimb.2010.02.091} {\path{doi:10.1016/j.nimb.2010.02.091}}.

\bibitem{david-bosne_generalized_2020}
E.~David-Bosne, U.~Wahl, J.~G. Correia, T.~A.~L. Lima, A.~Vantomme, L.~M.~C. Pereira, A generalized fitting tool for analysis of two-dimensional channeling patterns, Nucl. Instr. and Meth. B 462 (2020) 102--113.
\newblock \href {https://doi.org/10.1016/j.nimb.2019.10.029} {\path{doi:10.1016/j.nimb.2019.10.029}}.

\bibitem{wahl_position-sensitive_2004}
U.~Wahl, J.~G. Correia, A.~Czermak, S.~G. Jahn, P.~Jalocha, J.~G. Marques, A.~Rudge, F.~Schopper, J.~C. Soares, A.~Vantomme, P.~Weilhammer, Position-sensitive {Si} pad detectors for electron emission channeling experiments, Nucl. Instr. and Meth. A 524~(1) (2004) 245--256.
\newblock \href {https://doi.org/10.1016/j.nima.2003.12.044} {\path{doi:10.1016/j.nima.2003.12.044}}.

\bibitem{poikela_timepix3_2014}
T.~Poikela, J.~Plosila, T.~Westerlund, M.~Campbell, M.~D. Gaspari, X.~Llopart, V.~Gromov, R.~Kluit, M.~v. Beuzekom, F.~Zappon, V.~Zivkovic, C.~Brezina, K.~Desch, Y.~Fu, A.~Kruth, Timepix3: a {65K} channel hybrid pixel readout chip with simultaneous {ToA}/{ToT} and sparse readout, J. Inst. 9~(05) (2014) C05013.
\newblock \href {https://doi.org/10.1088/1748-0221/9/05/C05013} {\path{doi:10.1088/1748-0221/9/05/C05013}}.

\bibitem{burian_katherine_2017}
P.~Burian, P.~Broulím, M.~Jára, V.~Georgiev, B.~Bergmann, Katherine: {Ethernet} {Embedded} {Readout} {Interface} for {Timepix3}, J. Inst. 12~(11) (2017) C11001.
\newblock \href {https://doi.org/10.1088/1748-0221/12/11/C11001} {\path{doi:10.1088/1748-0221/12/11/C11001}}.

\bibitem{manek_track_2024}
P.~Mánek, P.~Burian, E.~David-Bosne, P.~Smolyanskiy, B.~Bergmann, Track {Lab}: extensible data acquisition software for fast pixel detectors, online analysis and automation, J. Inst. 19~(01) (2024) C01008.
\newblock \href {https://doi.org/10.1088/1748-0221/19/01/C01008} {\path{doi:10.1088/1748-0221/19/01/C01008}}.

\bibitem{wahl_direct_2020}
U.~Wahl, J.~G. Correia, R.~Villarreal, E.~Bourgeois, M.~Gulka, M.~Nesládek, A.~Vantomme, L.~M.~C. Pereira, Direct {Structural} {Identification} and {Quantification} of the {Split}-{Vacancy} {Configuration} for {Implanted} {Sn} in {Diamond}, Phys. Rev. Lett. 125~(4) (2020) 045301.
\newblock \href {https://doi.org/10.1103/PhysRevLett.125.045301} {\path{doi:10.1103/PhysRevLett.125.045301}}.

\bibitem{david-bosne_use_2021}
E.~David-Bosne, U.~Wahl, P.~A. Miranda, M.~Ribeiro~da Silva, E.~Alves, J.~G. Correia, Use of a {Timepix} position-sensitive detector for {Rutherford} backscattering spectrometry with channeling, Nucl. Instr. and Meth. B 499 (2021) 61--69.
\newblock \href {https://doi.org/10.1016/j.nimb.2021.05.005} {\path{doi:10.1016/j.nimb.2021.05.005}}.

\end{thebibliography}






\end{document}